\documentclass[3p,times,procedia]{elsarticle}
\usepackage{nupha_ecrc}

\volume{00}

\firstpage{1}

\journalname{Nuclear Physics A}

\runauth{Kolja Kauder \emph{for the JETSCAPE Collaboration}}

\jid{nupha}

\jnltitlelogo{Nuclear Physics A}

\usepackage{amssymb}

\usepackage{amsmath}

\usepackage{listings}
\usepackage[figuresright]{rotating}

\newcommand*{\xml}[1]{\texttt{<#1>}}

\begin{document}

\begin{frontmatter}



\dochead{}

\title{JETSCAPE v1.0 \\ Quickstart Guide}

\author{Kolja Kauder for the JETSCAPE Collaboration}
\address{Dept. of Physics, Brookhaven National Lab, Upton, New York 11973}
\ead{jetscape.org@gmail.com}
\ead[url]{http://jetscape.wayne.edu/}

\begin{abstract}
The JETSCAPE collaboration announced the first public release of its framework 
and Monte Carlo event generator at this conference, providing a unified interface and a comprehensive suite of 
model implementations for all stages of ultra-relativistic heavy ion collisions. 
This release focuses on validation of the framework and the $pp$ reference. 
A full manual is under development. In the mean-time, these proceedings will 
provide a guide for installation and simulation runs in lieu of the more traditional summary of the presentation.

\end{abstract}

\begin{keyword}
Heavy-Ion Collisions \sep Jet Energy Loss \sep Event Generator\sep Framework

\end{keyword}

\end{frontmatter}



\section{Introduction}
\label{introduction and Overview}
The JETSCAPE (Jet Energy-loss Tomography with a Statistically and Computationally Advanced Program Envelope)
collaboration was formed in 2016, as a joint effort of theoretical and experimental physicists, statisticians, and computer scientists,
with the mission of creating a complete, extensible, and modular event generator
using state-of-the-art computer science techniques in framework development,
as well as a complementary tool set for sophisticated statistical analysis (not contained in current release).
The event generator and the task-based framework with inter-task communication based on 
the signal-slot paradigm~\cite{external} and graph representation~\cite{external} of the shower structure 
are now publicly available.
While the $pp$ portion has been tuned, the heavy-ion portion requires calibration of several parameters, based on Bayesian
methods. This process, requiring  upwards of 30 million core hours, is currently being carried out. This uncalibrated (raw) version is being
released to seek usage data and community feedback. 

The currently included modules provide 
physics implementations for every stage of a collision (module name in parentheses):
TRENTO (\texttt{TrentoInitial}) for the initial state, a parton gun (\texttt{PGun}) as well as a PYTHIA8 interface (\texttt{PythiaGun}) for the initial hard scattering, 
simple brick (\texttt{Brick}) and Gubser (\texttt{GubserHydro}) hydrodynamics modules,
MATTER (\texttt{Matter}), MARTINI (\texttt{Martini}), Linear Boltzmann Transport
 (\texttt{LBT}), and AdS/CFT-based (\texttt{AdSCFT}) energy loss modules, 
as well as two PYTHIA-based hadronization modules (\texttt{Colored\-Hadronization, Colorless\-Hadronization}).
Additionally, wrappers exist that can be used for externally available packages, i.e.\ a pre-equilibrium free-streaming module
(\texttt{FreestreamMilneWrapper}), viscous hydrodynamics with MUSIC (\texttt{MpiMusic}), 
and a freeze-out surface sampler (\texttt{iSpectraSamplerWrapper}), with download scripts
available in the \texttt{external\char`_packages} directory.
Initial state and hydro history can also be read from suitably formatted HDF5 files~\cite{external},
please contact the collaboration for details.
References and license information can be found in the \texttt{COPYING} and AUTHORS files.

\section{Installation}
\label{sec:installation}
The package can be downloaded from the official repository on GitHub~\cite{jsrepo}.
An existing \texttt{PYTHIA8}~\cite{Sjostrand:2007gs} installation is assumed.
The framework is designed to rely only on minimal prerequisites, but some included and
optional physics modules require additional packages, most notably the \texttt{Boost} libraries~\cite{external};
for a full list as well as more detailed installation instructions, please refer to the README files. 
After downloading, please create and descend into a \texttt{build} directory where
you can configure with cmake~\cite{external}, and compile:
\begin{lstlisting}
  $ cmake .. &&  make
\end{lstlisting}

A few options, such as HepMC, are automatically recognized during configuration, 
but some modules have to be activated using cmake switches.
The following example will (after additional downloads) make available the MUSIC module as well as a free streaming and a surface sampling module:
\begin{lstlisting}
  $ cmake -Dmusic=on -DiSS=on -Dfreestream=on ..
\end{lstlisting}

\section{Validated $pp$ Reference}
\label{sec:pp-reference}

\begin{figure}
  \centering
  \includegraphics[width=0.45\textwidth]{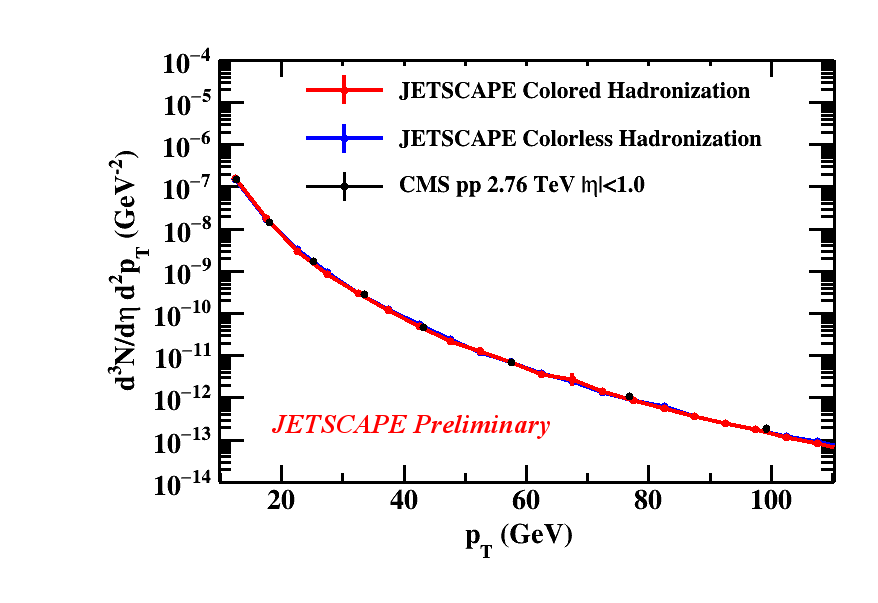}
  \includegraphics[width=0.45\textwidth]{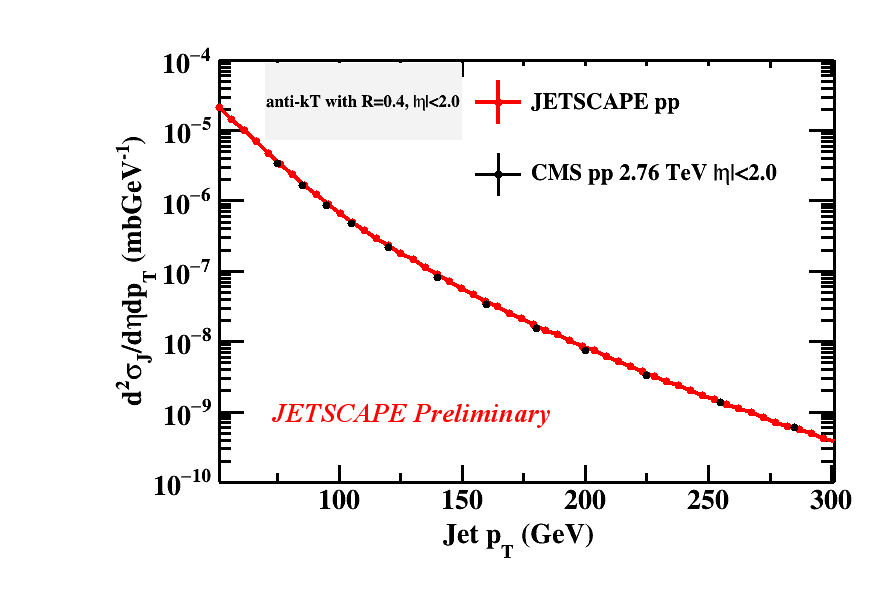}
  \caption{Charged hadron spectrum (left) and $R=0.4$ anti-$k_T$ jet spectrum (right) compared to $pp$ data at $\sqrt{s}=2.76$~TeV.}
  \label{fig:spectra}
\end{figure}

The mandate of the JETSCAPE collaboration is to release the code to the community as soon as it is ready for heavy-ion analyses.
As such the $pp$ portion of the code has been tuned to reproduce most of the jet data at the level of agreement displayed by PYTHIA.
Calibration of the heavy-ion sector is currently being carried out. 
Figure~\ref{fig:spectra} demonstrates excellent agreement with published data~\cite{CMSboth}. 
The following is a guide through the steps to recreate these figures.

The JETSCAPE event generator is steered in two places, a wrapper program and an XML configuration file.
To create the $pp$ reference, the initial hard scattering is created 
with PYTHIA8 (with multi-parton interaction (MPI) and initial state radiation (ISR) turned on), 
final state radiation (FSR)  is handled by MATTER~\cite{matter} with $\hat q$ set to 0,
and hadronization is again done by PYTHIA in one of two available modes, i.~e. with and without preserving color information.
The wrapper used for $pp$ runs is found in \texttt{examples/PythiaBrickTest.cc}. 
The following line initializes JETSCAPE with an XML file and 200 events:
\begin{lstlisting}
  auto jetscape = make_shared<JetScape>("./jetscape_init.xml",200);
\end{lstlisting}
Note that the default file (\texttt{jetscape\char`_init.xml}) in the build directory is overwritten 
by subsequent calls to cmake, and thus it is highly recommended to use a different filename. 
Physics modules are attached using a shared \emph{auto} pointer that resolves to the required type:
\begin{lstlisting}
  auto pythiaGun= make_shared<PythiaGun> ();  jetscape->Add(pythiaGun);
  auto hydro = make_shared<Brick> ();  jetscape->Add(hydro);
\end{lstlisting}
Energy loss (or vacuum radiation in the case of MATTER with $\hat q=0$) 
and hadronization are controlled by \emph{managers} and need a few extra lines.
The following example demonstrates energy loss; the code is similar for hadronization:
\begin{lstlisting}
  auto jlossmanager = make_shared<JetEnergyLossManager> ();
  auto jloss = make_shared<JetEnergyLoss> ();
  auto matter = make_shared<Matter> ();   jloss->Add(matter);
  jlossmanager->Add(jloss);    jetscape->Add(jlossmanager);
\end{lstlisting}
Finally, to create output files in the custom JETSCAPE format, use:
\begin{lstlisting}
  auto writergz= make_shared<JetScapeWriterAsciiGZ> ("test_out.dat.gz");
\end{lstlisting}
Without the ``gz'', the created output is immediately human readable, 
but for large-scale productions we strongly recommend this space-saving 
gzipped alternative. A provided reader can accomodate either choice without the need 
to manually unzip the output.
Alternatively, a writer for HepMC3~\cite{external} is provided for users more familiar with this format.

The configuration file controls settings for all modules in XML format. 
For the purpose of running in $pp$ mode, only the following tags are relevant:
\begin{description}
\item[\xml{Hard} $\rightarrow$ \xml{PythiaGun}:]
  \xml{eCM} controls the center of mass energy in GeV, 
  \xml{pTHatMin} and \xml{pTHatMax} correspond to PYTHIA's \texttt{PhaseSpace:pTHat} settings. 
  In order to scan the desired parameter space in discrete bins of \texttt{pTHat}, one should use
  multiple configuration files. Appropriate weights are saved in the output file for later recombination.
\item[\xml{Eloss} $\rightarrow$ \xml{Matter}:] Parameters are tuned, just set \xml{qhat0} to 0.0 and \xml{in\char`_vac} to 1.
\item[\xml{JetHadronization}:] PYTHIA maintains color connections throughout the entire event;
  this information is only partially retained within JETSCAPE.
  For colored hadronization, the color of the initiating parton and each parton in the ensuing shower is retained.
  A beam remnant with a specific color at large rapidity is added to each shower to make it a color singlet;
  the \xml{eCMforHadronization} tag
  controls the energy of artificially added beam remnant proxies.
  There is some freedom for this parameter, we recommend $\sqrt{s}/2$.
\item[\xml{Random}:] Optionally, a random \texttt{seed} can be specified here to identically recreate prior execution. 
  As long as all modules use the provided \texttt{GetMt19937Generator()} functionality, results will be reproducible throughout the framework.
  For batch execution, this parameter should be unique for every job, alternatively it can be set to 0 which will choose 
  a seed based on the current date and time.
\end{description}

A technical side note: Internally, the \texttt{PythiaGun} module initiates individual showers for all particles with status code 62 to
account for MPI. If you wish to change this behavior, to for example only use the two hardest partons (status -23) when turning off MPI,
it will be necessary to modify, or better yet rename and modify, the physics module itself at \texttt{initialstate/Pythiagun.cc}.

After configuration and compilation, execute the wrapper with the following command:
\begin{lstlisting}
  $ ./bin/PythiaGun
\end{lstlisting}
For practical purposes, it may be advisable to modify this wrapper to 
accept command line arguments and be more amenable to running the necessary jobs 
for all desired \texttt{pTHat} bins within a given cluster farm.

Two examples are provided to process the created output.
The simplest way is shown in \linebreak[5]
 \texttt{examples/FinalStateHadrons.cc} which extracts final state hadrons
from the full shower and writes them into a text file for further analysis with a chosen histogramming 
package.
A more advanced example can be found in \texttt{examples/readerTest.cc} where FastJet is directly run on the extracted particles. 
In this example, we only use the included lightweight \texttt{fjcore} library, but the full FastJet package and other software such as ROOT
can also be used. Note that \texttt{cmake} automatically recognizes an existing ROOT installation, and the top level file \texttt{CMakeLists.txt}
contains commented-out examples that demonstrate how to link against these libraries.
For weighting according to \texttt{pTHat}, the generated cross section (and its uncertainty) is saved for every event.
This value updates throughout the generation process, so it is recommended to only use the last one which can be obtained 
from the output file for example via:
\begin{lstlisting}
  $ zgrep sigma test_out.dat.gz|tail -n 2
  # sigmaGen 8.35147e-06
  # sigmaErr 3.2343e-07
\end{lstlisting}

\section{Next Steps and Outlook}
\label{sec:outlook}

Users will naturally want to explore energy loss within a realistic expanding medium next,
and all provided physics modules in the current release are fully functional to do just that.
Examples provided in \texttt{examples/MUSICTest.cc} and \texttt{examples/freestream-milneTest.cc}
show how to interface with the (3+1)-dimensional viscous hydrodynamics from MUSIC~\cite{music}
(currently only in 2+1 dimensions), the former with an additional freeze-out surface sampler to
create complete events comprised of bulk and shower hadrons, the latter with a free-streaming
module between initial state and hydro evolution. 

Also of note is the ability to combine multiple energy loss modules, for example by 
adding   
\begin{lstlisting}
   auto martini = make_shared<Martini> ();   jloss->Add(martini);
\end{lstlisting}
to any example wrapper. 
The use of multiple modules requires the user to specify well defined boundaries,
so that a given parton at one space-time point is not acted upon by multiple energy loss modules.
The prescribed method for MATTER+MARTINI or MATTER+LBT is to use 
parton virtuality and a switching point at $1.0~\text{GeV}/c^2$.
This feature of multi-stage energy loss~\cite{Majumder:2010qh,Cao:2017zih}, where 
distinct models are responsible only in their region of applicability,
and the ability to add more modules and modify their boundaries remains one of the unique features of the JETSCAPE framework.
the implementation of unambiguous boundaries is left to the discretion of the user.
A framework-level implementation that is at the same time user-friendly and flexible enough
to handle arbitrarily-shaped multi-dimensional parameter spaces to allow for example 
switching on virtuality, energy and surrounding medium temperature
is a very challenging task which will require additional development.

Upcoming releases in the near future will focus in two directions:
On the one hand, incorporation of additional physics
such as hadronic rescattering with SMASH~\cite{Weil:2016fxr} and medium recoil are a high priority.
On the other hand, the existing modules are in the process of being tuned both in regards
of the interplay between initial stage and hydrodynamic evolution 
and parameters of energy loss models as well as the switching between them.

JETSCAPE is community software: As such we are are appreciative of, and crucially dependent on, any and all feedback from the community,
from installation and documentation issues to bug reports and suggestions. 
Bugs and feature requests are best reported directly in the GitHub issue tracker~\cite{jsrepo},
general questions and suggestions should be sent to the email provided in the contact information.
\\
\\
These proceedings are supported by the US NSF under the grant \# 1550300.




\bibliographystyle{elsarticle-num}
\bibliography{ref}







\end{document}